\documentclass{jnmp}

\usepackage{amsmath}
\JNMPnumberwithin{equation}{section}

\newtheorem{proposition}{Proposition}[section]

\theoremstyle{definition}

\newtheorem{example}{Example}[section] % The '*' makes it unnumbered

\begin{document}

%  Headings
%
\renewcommand{\evenhead}{I~A~Shereshevskii}
\renewcommand{\oddhead}{A Finite Dimensional Analog of the Krein Formula}

%  Titlepage
%
\thispagestyle{empty}

\FirstPageHead{8}{4}{2001}{\pageref{sheresh-firstpage}--\pageref{sheresh-lastpage}}{Letter}
%  Parameters: Volume, number, year, page range, paper type
%  'Article' could be changed to 'Letter' or 'Review Article'

\copyrightnote{2001}{I~A~Shereshevskii}

\Name{A Finite Dimensional Analog\\ of the Krein Formula}
\label{sheresh-firstpage}

\Author{I~A~SHERESHEVSKII}

\Address{Institute for Physics of Microstructures,  Russian Academy of Sciences, \\
 GSP-105,  Nizhny Novgorod,  RU-603950, Russia\\
E-mail: ilya@ipm.sci-nnov.ru}

\Date{Received September 19, 2000; Revised June 29, 2001; Accepted
July 3, 2001}

\begin{abstract}
\noindent
I offer a simple and useful formula for the resolvent
of a small rank perturbation of large matrices.  I discuss
applications of this formula, in particular, to analytical and
numerical solving of difference boundary value problems.  I present
examples connected with such problems for the difference Laplacian and
estimate numerical efficiency of the corresponding algorithms.
\end{abstract}

\section{Introduction}\label{Int}

Wide application of various versions of M~G~Krein's formula resulted
lately in a marked progress of the theory of boundary value problems
for equations of mathematical physics.  In its initial form, this
formula connects the resolvents of two different selfadjoint
extensions of a given symmetric operator with finite defect indices in
the Hilbert space \cite{AG}.  Using this connection one can find, in
particular, the exact solution of the Schr\"{o}dinger equation with
the point-wise potentials, construct the correct theory of the
boundary problems for the Laplace operator on graphs \cite{Graph,
Graph2}, obtain an approximate expression for the resolvent and the
exponent of the second order differential operator in the domains of
${\mathbb  R}^{n}$ in terms of the parametrix of the operator in the whole
space \cite{Pav, Ant}, etc.  The formula for the resolvent of the
boundary value problems, which appears in such a way, can also be used
for the construction of numerical algorithms.

It is interesting to find a direct analog of M~G~Krein's formula for
the difference equations (and here such an analog is offered).  This
is important for construction of numerical algorithms when we reduce
the initial differential equation to some finite dimensional (i.e.,
matrix) problem.  If we had a discrete analog of M~G~Krein's
formula, we would have been able to develop effective numerical
algorithms for solution of the difference boundary value problems.

I begin with a very simple matrix relation.  For reasons which I will
try to explain in what follows, I refer to it as {\it finite
dimensional analog of the Krein formula}.  After brief discussion, I
present a few examples, showing the simplest applications of this
relation.  In particular, I give a short description of the algorithm
for solving general boundary value problem for difference Laplacian in
two-dimensional rectangular domain with (perhaps) small defects such
as holes or cuts.

\section{Low dimensional perturbations of the matrix\\
and their resolvents}\label{KMR}

We start with a very simple question from Linear
Algebra.  Though rather important, as I will try to illustrate, it did
not attract attention of researchers, at least, I could not find it in
the literature.  Even Prasolov's encyclopedia of nice problems in
Linear Algebra~\cite{P} missed it.  Namely, suppose $A$ is a square
matrix and we had computed its inverse $A^{-1}$.  And --- such a
terrible but common disaster!  --- we observe that the typist has
typed {\it one} of matrix elements of the initial matrix wrong!  Must
we redo the whole work (this is pretty expensive for large matrices!)
or there is a cheaper possibility to obtain the correct answer?

I will show an almost obvious way to answer the last half of the
question in affirmative: there is a cheaper way.  I assume, of course,
that all the matrices we are going to invert are indeed invertible.

Thus,  suppose we have to solve the equation
\begin{equation}
    (A+B)x=f,    \label{init}
\end{equation}
where $x, f\in {\mathbb  C}^{n}$, $f$ is a known vector and $x$ is an
unknown one.  If we can easily solve the ``unperturbed'' equation
$Ax=g$ for any right hand side $g$ (this means exactly that $A^{-1}$
is known), we can re-write (\ref{init}) in the form
\begin{equation}
    \left(E+A^{-1}B\right)x=A^{-1}f.
    \label{st1}
\end{equation}
We introduce new unknown vector $z$ from the relation $z=Bx$.
Then,  if we multiply (\ref{st1}) by $B$ from the left,  we obtain the
equation for $z$:
\begin{equation}
    \left(E+BA^{-1}\right)z=BA^{-1}f.
    \label{st2}
\end{equation}
The following evident assertion  holds.

\begin{proposition}
If both matrices $A$ and $A+B$ are
invertible,  then $E+BA^{-1}$ is also invertible.
\end{proposition}

\begin{proof} Assume the contrary. This means that the homogeneous equation
 \begin{equation}
    \left(E+BA^{-1}\right)z=0
    \label{interm}
\end{equation}
has a nontrivial solution $z_{0}\neq 0$.  Since $A$ is invertible
by hypothesis,   the vector $x_{0}=A^{-1}z_{0}$ exists and is nonzero.  Then
having substituted $z_{0}=Ax_{0}$ into (\ref{interm}) we
obtain
\[
      (A+B) x_{0}=0,
\]
in contradiction with the fact that $A+B$ is invertible.
\end{proof}

We can,  therefore,  solve equation (\ref{st2}) and write
\begin{equation}
    z= \left(E+BA^{-1}\right)^{-1}BA^{-1}f.
    \label{st3}
\end{equation}
It is clear that equation (\ref{st1}) can be now rewritten as
\begin{equation}
       x=A^{-1}f-A^{-1}z.
    \label{st4}
\end{equation}
Now,  we can substitute expression (\ref{st3}) for vector $z$ into
(\ref{st4}) and finally obtain the solution of (\ref{init}) in the
following strange form:
\begin{equation}
    x= A^{-1}f-A^{-1}\left(E+BA^{-1}\right)^{-1}BA^{-1}f.
    \label{vKrein}
\end{equation}
We can also rewrite this formula as an operator relation:
\begin{equation}
\fbox{$(A+B)^{-1}= A^{-1}-A^{-1}\left(E+BA^{-1}\right)^{-1}BA^{-1}$}
    \label{Krein}
\end{equation}
which I refer in what follows as the {\it finite dimensional analog of
Krein's formula.}

Obviously,  all the above is meaningless for {\it generic} matrices $A$
and $B$,  because the calculation of the inverse matrix for $E+BA^{-1}$
is of the same complexity as that of the initial one.  The situation
changes dramatically if the {\it rank of $B$ is small}.

The word ``small'' means in this context that the ratio
$\frac{\mbox{rk}\,B}{\mbox{rk}\, A}$ is much smaller than $1$.  In this
case the calculation of the inverse matrix of $E+BA^{-1}$ becomes
simple.

\begin{example}  Let $V$ be an $n$-dimensional vector space,
$e\in V$ and $f\in V^{*}$.  Let $B=e\otimes f$ be the linear operator
in $V$ of rank $1$.  Then it is easy to see that
\begin{equation}
    \left(E+BA^{-1}\right)^{-1}=E-\left(1+f\left(A^{-1}e\right)\right)^{-1}e\otimes\left(A^{-1}\right)^{*}f,
    \label{Ex1}
\end{equation}
and,  therefore,  we need only about $n^{2}$ operations to calculate
matrix $\left(E+BA^{-1}\right)^{-1}$ instead of about $n^{3}$ in the general
case.

One can obtain the estimate $m^{3}+mn^{2}$ for complexity of such a
calculation when $\mbox{rk}\,B=m$.  This value is much smaller than
$n^{3}$ provided $\frac{m}{n}\ll 1$.  So we can consider relation
(\ref{Krein}) as a version of perturbation theory in which the ratio
$\frac{\mbox{rk}\,B}{\mbox{rk}\,A}$ of the rank of perturbation to the
rank of the unperturbed operator plays the role of small parameter.

We will see in what follows that in some important cases the
calculation of the unperturbed resolvent may turn out to be incredibly
simple,  in distinction with direct calculation of the perturbed one,
and in these cases application of formula (\ref{Krein}) becomes very
effective.
\end{example}

Before demonstrating possible applications of formula
(\ref{Krein}) in computational mathematics,  let me briefly explain the
reason to baptize a very simple relation from Linear Algebra with a
famous name.  As it is mentioned in Introduction,  the ``actual'' Krein
formula \cite{AG} connects the resolvents of two different
self-adjoint extensions $A_{1}$ and $A_{2}$ of a given symmetric
operator $A_{0}$ in an {\it infinite dimensional} Hilbert space $H$.

Unfortunately, it is very difficult, or, perhaps, even impossible, to
read any operator sense into the difference $A_{1}- A_{2}$ of such
extensions, because, as a rule, this difference vanishes on the
intersection of their domains.  Such and similar difficulties,
however, had never been an obstacle for physicists, and they eagerly
used Dirac's $\delta$-function as a potential of ``point-wise
interaction'' in the Schr\"{o}dinger equation, see, e.g., \cite{Alb}.
Certain arguments which I skip convinced me that ``point-wise''
perturbations of differential operators are in some sense {\it
perturbations of finite rank} and, due to this fact, the corresponding
problems have exact solutions.

Note in this connection that, although the actual Krein formula
presupposes finiteness of defect indices of the initial symmetric
operator, it may by used as well in the case of infinite indices
(there are a number of papers on this topic, see e.g., \cite{AlbKur,
KurKur} and references therein).  Such a case arises, e.g., if we
consider different boundary value problems for given symmetric partial
differential operator \cite{Pav}.

In what follows we consider a problem of calculating resolvents of
extensions (see formal definition in the next section) of {\it
difference} operators based on formula (\ref{Krein}).

It seems that the corresponding relation is a finite dimensional
analog of the relation for differential operators.

This impression is not an illusion.  Indeed, it is possible to
consider (in some well-defined sense) the difference operators as
approximation of differential ones, and then one can prove that {\it
in the case of finite defects} (e.g., for ordinary differential
operators) our formulas converge to the corresponding formulas for
differential operators (private communication of E~Gordon and
S~Albeverio; together with them we intend to explain this in detail
elsewhere).  Such convergence plays a crucial role both for goals of
numerical analysis and as an instrument for investigation of infinite
dimensional operators via their finite dimensional approximations
(see, e.g., \cite{Gor}).  Unfortunately, rigorous results about
convergence of finite-dimensional approximations of operators requires
for proofs a nonelementary technique which is out of frame of this
work.  Nevertheless, I consider (briefly and without proof) at the end
of Section~\ref{BVP} the simplest example of convergence of finite
dimensional Krein formula for difference approximations of the
Schr\"odinger operator with $\delta$-potential on the unit circle to
the usual Krein's formula for the resolvent of this operator.

\section{Boundary-value problem for the difference
operators}\label{BVP}

Difference approximations of boundary value problems for
differential operators are a base for the numerical solving of such
problems.  Here I just introduce a convenient for our nearest goals
language for formal description of ``abstract'' difference boundary
value problems.  I could not find an appropriate analog of such a
language in the literature.  Hopefully, the following examples make it
clear why this language is useful and convenient.

For any set ${\mathcal M}$,  let $C({\mathcal M})$ be the space of all
complex-valued functions on ${\mathcal M}$.

A linear map $ A : C({\mathcal M})\longrightarrow C({\mathcal M})$ will be
called a {\it formal difference operator} in $C({\mathcal M})$ if for each
$x\in {\mathcal M}$ there exist a {\it finite} set $\gamma_{A}(x)\subset
{\mathcal M}$ and function $a_{x}\in C(\gamma_{A}(x))$ such that
\begin{equation}
( Af)(x)=\mathop{\sum}\limits_{y\in \gamma_A (x)}
a_{_{x}}(y) f(y),  \qquad f\in C({\mathcal M}),  \label {dcop}
\end{equation}

Let $\Omega $ be a subset of ${\mathcal M}$.  The point $z \in \Omega $ is
an {\it inner point of the set $\Omega$ with respect to the map} $ A$,
if $\gamma_A(z) \subseteq \Omega$.

The point $z \in \Omega $ is a {\it boundary point of the set $\Omega$
with respect to the map} $A$,  if $\gamma_A(z)\setminus \Omega\neq
\emptyset$.  The {\it boundary of $\Omega$ with respect to } $ A$ is
the set $\partial _{A}\Omega$ of all boundary points of $\Omega$.
Define the set $b_{A}\Omega$ of {\it exterior points of $\Omega$ with
respect to} $ A$ to be
\[
b_{A}\Omega = \bigcup_{x\in \partial_{A} \Omega} (\gamma_A(x) \cap
\overline{\Omega}),  \qquad \mbox{where}\quad \overline{\Omega}= {\mathcal M}\setminus
\Omega.
\]
Note that in the ``difference'' case the sets $\partial _{A}\Omega$
and $b_{A}\Omega $ do indeed depend on the map $ A$ in contrast with
the continuous situation.  Observe that
\[
\bigcup_{ x \in \Omega } \gamma_A(x) = \Omega \cup b_{A}\Omega .
\]
Define the map $ A_{\Omega} : C(\Omega \cup b_{A}\Omega)
\longrightarrow C(\Omega)$ by formula (\ref{dcop}) for any point $x
\in \Omega$.

Let $ L:  C(\Omega) \longrightarrow C(\Omega \mathop{\cup}\limits
b_{A}\Omega)$ be a linear map such that $( Lf)(x)=f(x)$ for all $x \in
\Omega$.  The operator $L$ will be called an {\it extension operator}
for the map $A$.

We say that the operator $A_{L}: C(\Omega) \longrightarrow C(\Omega)$
is an $L$-extension of the formal difference operator $A$ if
\begin{equation}
( A_{L} f)(x)=( A_{\Omega} L f)(x),  \qquad x \in \Omega.
\label {lexpa}
\end {equation}

Note,  that the in case described,  the extension operators play the
role of boundary conditions for differential operators.  I hope that
this will be clear from the examples of this section.

In what follows I suppose that the set $\Omega$ is finite.
Let for $\lambda \in {\mathbb C}$ the operator $R_{L}(\lambda )$ be
the resolvent of the $ L$-extension of the operator $A$ such that
$R_{L}(\lambda ) = \left( A_{L}- \lambda E\right)^{-1}$.

We show that formula (\ref{Krein}) establishes a simple algebraic
connection between resolvents $R_{L}(\lambda )$ and $R_{K}(\lambda )$
of two different extensions of the formal difference operator $ A$
corresponding to two extension operators $ L$ and $ K$.  (In what
follows I assume that $\lambda$ is a common resolvent point for both
$A_{L}$ and $A_{K}$.)  To obtain such a connection,  note first that
definition (\ref{lexpa}) implies
\begin {equation}
A_{K} = A_{L} +D_{LK} ,  \qquad \mbox{where} \quad D_{LK}= A_{\Omega}( K- L).  \label {sumop}
\end {equation}
Now,  let us replace matrix $A^{-1}$ in (\ref{Krein})
with $ R_{L}(\lambda )$,  matrix $(A+B)^{-1}$ with $ R_{K}(\lambda
)$,  and $B$ with $D_{LK}$.  Then we see that
\begin {equation}
R_{K}(\lambda)= R_{L}(\lambda )- R_{L}(\lambda )(E+D_{LK}R_{L}(\lambda
) )^{-1} D_{LK} R_{L}(\lambda ).  \label {dicKre}
\end {equation}

Observe that all the inverse operators in this formula exist by the
hypothesis.

{\bf What do we gain from this formula?}  Note first of all,
that it is easy to see that
\[
\mbox{rk}\, D_{LK}\leq \# (\partial_{A}\Omega),  \qquad \mbox{and}
\qquad \mbox{rk}\, (A_{K}-\lambda)=\mbox{rk}\, (A_{L}-\lambda E)=\# (\Omega).
\]
It is remarkable that as a rule (see examples in what follows) $\#
(\partial_{A}\Omega)\ll \# (\Omega)$,  and we are in the situation
discussed in Section~\ref{Int}.  Examples also show that the complexity
of calculation of the resolvent for different extensions may be
essentially different.

\begin{example}
{\bf Resolvent of the one-dimensional difference
Laplacian.} Although it seems that this example has no practical meaning,  it makes
very clear all previous abstract constructions and has all essential
features of practically important Example 3.2.

Let ${\mathcal M}={\mathbb  Z}$ the set of integers,  $N$ a positive integer,
and $\Omega = \{ 0,  1,  \ldots ,  N-1\}$.  The one-dimensional difference
Laplacian is the formal difference operator $\Delta : C({\mathcal M})
\longrightarrow C({\mathcal M})$ defined by the relation
\[
(\Delta f)(x) = f(x+1) -2 f(x) + f(x-1),  \qquad \mbox{where} \quad  x \in {\mathcal M}
\quad \mbox{and} \quad  f \in C({\mathcal M}).
\]
Note that operator $\Delta$ differs from the usual difference
approximation of the differential expression $\frac{d^{2}}{dx^{2}}$
on the uniform grid in ${\mathbb R}^{1}$ by a factor only.

In the case considered $\partial _{\Delta} \Omega=\{ 0,  N-1\}$ and
$b_{\Delta} \Omega=\{-1,  N\}$.

We list all the extensions for $\Delta$.  Let $\hat{l}:
C(\Omega)\longrightarrow C(\{-1,  N\})$,  i.e.,  $\hat{l}$ is represented
by a $2\times N$ complex matrix.  Then for $f\in C(\Omega)$ set
\[
  (Lf)(x)=f(x),  \quad x\in\Omega,  \qquad (Lf)(y)=(\hat{l}f)(y),  \quad y\in
  b_{\Delta}\Omega.
\]
Clearly,  any extension operator for $\Delta$ must be of such form.

Among all $L$-extensions of $\Delta$ there exists an exceptional one,
for which the corresponding resolvent has an ``almost explicit''
expression.  This is the so-called {\it periodic extension},  defined
by extension operator $L_{0}$ such that
\begin{equation}\label{PBC}
 \left(\hat{l}_{0}f\right)(-1)=f(N-1),   \qquad  \left(\hat{l}_{0}f\right)(N)=f(0).
\end{equation}
The exceptional role of this extension (denoted in what follows by
$\Delta_{0}$ instead of $\Delta_{L_0}$ for brevity) is the consequence
of the fact that it can be diagonalized by Discrete Fourier
Transformation (DFT),   i.e.,
\begin {equation}
\Delta_{0}=F\Lambda F^{*},
\label {furlap}
\end {equation}
where $\Lambda$ is the
multiplication operator (i.e.,   the diagonal matrix)
\[
(\Lambda f)(x)=-4\sin^{2}\frac{\pi x}{N}f(x),   \qquad  x\in\Omega,
\]
and the unitary  DFT operator $F$ is defined by the relation
\[
(Ff)(x) =\frac{1}{\sqrt{N}}\sum_{y\in \Omega}e^{-i\frac{2\pi xy}{N}}f(y).
\]
Formula (\ref{furlap}) immediately implies the equality
\begin{equation}
R_{0}(\lambda )=F(\Lambda- \lambda E)^{-1}F^{*},
\label{res0}
\end {equation}
and this is what we meant under the {\it explicit formula for resolvent}.

It is well known that there exists an abnormally effective numerical
method (called Fast Fourier Transformation,  or FFT) for application of
DFT to the vector.  It requires only $\sim N\log N$ arithmetic
operations instead of $\sim N^2$ for the general $N\times N$ matrices
\cite{Bah}.  This fact crucially reduces the complexity of computation
of operator (\ref{res0}).

Is there an algorithm which allows one to calculate the resolvent of an
{\it arbitrary} extension of $\Delta$ with the same complexity as for
$\Delta_{0}$?  Formula (\ref{dicKre}) gives a positive answer to this
question.  It only suffices to show that the computation of matrix
$(E+D_{0K}R_{0}(\lambda))^{-1}$ is not a problem.  Indeed,  due to the
fact that $(D_{0K}f)(x)=0$ for $f\in C(\Omega)$ and $x\in
\Omega\setminus \partial_{\Delta}\Omega$,  to solve the equation
\begin{equation}
 (E+D_{0K}R_{0}(\lambda))f=g,  \label {boueq}
\end {equation}
we only have to find $f(0)$ and $f(N-1)$.  We denote by
$\delta_{x}$ the function from $C(\Omega)$ defined by
\[
\delta_{x}(y)=\left\{\begin {array}{ll}
0,  & \mbox{for}\ x\neq y,  \\
1,   & \mbox{for}\  x=y.
\end{array}\right.
\]
Since $f(x)=g(x)$ for $x\in \Omega\setminus
\partial_{\Delta}\Omega$,   we can re-write  equation (\ref{boueq}) in
the form
\begin{gather*}
(1+(D_{0K}R_{0}(\lambda)\delta_{0})(0))f(0)+
         (D_{0K}R_{0}(\lambda)\delta_{N-1})(0)f(N-1)\\
\qquad =g(0)-(D_{0K}R_{0}(\lambda)(g-g(0)\delta_{0}-g(N-1)\delta_{N-1}))(0),\\
(D_{0K}R_{0}(\lambda)\delta_{0})(N-1)f(0)+
      (1+(D_{0K}R_{0}(\lambda)\delta_{N-1})(N-1))f(N-1)\\
\qquad =g(N-1)-(D_{0K}R_{0}(\lambda)(g-g(0)\delta_{0}-g(N-1)\delta_{N-1}))(N-1).
\end{gather*}
This is a system of two linear equations for two unknowns,  which is
solvable due to Proposition~2.1.  So,  to calculate the resolvent
$R_{K}(\lambda)$,  we only have to know how to calculate
$R_{0}(\lambda)$ and how to invert $2\times 2$-matrices \ldots
\end{example}

This example is,  as have already been said,  of no practical
importance,  because there exists another (not DFT-based) algorithm for
inverting the general three-diagonal matrix of complexity $\sim N$ (so
called {\it sweep method},  see,  e.g.,  \cite{Bah}).  For most often
used types of boundary conditions (i.e.,  extension operators $K$),  the
matrix of $\Delta_{K}$ is of this kind,  and the sweep method becomes
preferable.  For example,  the Dirichlet problem corresponds to the
extension defined by the map $\hat{l}$ of the form
\[
\left(\hat{l}f\right)(-1)=-f(0),   \qquad \left(\hat{l}f\right)(N)=-f(N-1).
\]
and this leads to a three-diagonal matrix.

Note,  however,  that if for an extension operator $K$ the matrix of
$\Delta_{K}$ is not three-diagonal (as is the case,  e.g.,  for
$\Delta_{0}$),  one can use the Dirichlet extension as the ``initial''
one and solve the problem for the $K$-extension using only $\sim N$
arithmetic operations!  The reason for using DFT in this example
becomes clear from the following example.

\begin{example}
{\bf The boundary value problem of third kind
for Laplacian in two-dimensional rectangle.}
We consider now the boundary value problems for the
two-dimensional difference Laplacian.  Let ${\mathcal M}={\mathbb  Z}^{2}$,  $N$
and $M$ positive integers,  and $\Omega = \{0,  \ldots ,  N-1\} \times
\{0,  \ldots ,  M-1\}$.  The formal two-dimensional Laplace operator
which we denote by the same symbol $\Delta: C({\mathcal M})\longrightarrow
C({\mathcal M})$ is given by the formula:
\begin{gather*}
    (\Delta f)(x,   y) = f(x+1,   y) + f(x-1,   y)+ f(x,   y+1) + f(x,   y-1) -
4f(x,   y),  \\
\mbox{for} \quad x,   y \in {\mathcal M} \quad \mbox{and}  \quad f \in C({\mathcal M}).
\end{gather*}
Clearly,
the set of the boundary points with respect to the operator $\Delta$
is
\[
\partial _{\Delta } \Omega=
\left (\{0,  N-1\}\times \{0,  \ldots ,   M-1\}\right ) \cup
\left (\{0,  \ldots ,  N-1\}\times  \{0,  M-1\}\right ),
\]
so that $\# (\partial _{\Delta } \Omega)=2(N+M-2)$.  We see once
more that $\# (\partial _{\Delta } \Omega )\ll \#
(\Omega)=MN$.  Hence,  there exists a good chance for applying Krein's
formula.  To actually apply it,  we first describe the set
$b_{\Delta }\Omega$.
\end{example}

The next geometric proposition is almost evident
and we omit proof.

\begin{proposition}
 1) For $p=(x,  y)\in {\mathbb  Z}^{2} $ set
$|p|=|x|+|y|$.  Then $p\in b_{\Delta }\Omega $ if and only if there
exists (and then it is unique) $\varepsilon(p)\in {\mathbb  Z}^{2}$ such
that $|\varepsilon(p)|=1$ and $p+\varepsilon(p)\in \Omega$.

2) $\# (b_{\Delta }\Omega) =2N+2M$.
\end{proposition}

We will not describe all extension operators for $\Delta$ (though
possible,  this is not interesting),  instead we will consider several
distinguished cases.  First of all,  as in the one-dimensional case
considered in Example 3.1,  there exists a remarkable periodic
extension defined by extension operator $L$ of the form
\[
   (Lf)(x,  y)=f(x\ {\rm mod}\; N, \ y\ {\rm mod}\;  M),
     \qquad   (x,  y)\in b_{\Delta }\Omega.
\]
The corresponding operator will be denoted again by $\Delta _{0}$ and
it has the same characteristic property,  namely,  may be diagonalized
by a {\it two-dimensional} DFT \cite {Bah}.  Therefore,  one needs
$\sim MN\log MN$ arithmetic operations for calculating the resolvent
$R_{0}(\lambda)$ instead of about $(MN)^{3}$ to invert the general
linear operator in $C(\Omega)$.

Among other extensions of two-dimensional difference Laplacian,  I
consider only the ones corresponding to {\it local boundary
conditions} for the differential Laplace operator.  These extensions
are defined by the family of extension operators $K$ of the form
\begin {equation}
(Kf)(p)=k(p)f(p+\varepsilon(p)),  \qquad p\in
b_{\Delta }\Omega,
   \label{bouloc}
\end {equation}
where $k\in C(b_{\Delta }\Omega)$ and
$\varepsilon(p)$ is defined in Proposition 3.1.

We now consider again the relation (\ref{boueq}).  It is easy to
see that,  as in the one-dimensional case,  this equation can be
transformed to a linear equation for function $f\in C(\partial_{\Delta
}\Omega)$ and we need $\sim (2M+2N-2)^{3}$ arithmetic operations to
solve it.  For $M$, $N$ large enough,  the inequality $(2M+2N-2)^{3}\ll
(MN)^{3}$ holds,  and we obtain the algorithm for solving the third
kind boundary value problem for two-dimensional difference Laplacian
with complexity $\sim (2M+2N-2)^{3}+MN\log MN$ arithmetic operations.
Moreover,  if one has to repeatedly solve this problem for different
right hand sides,  it suffices to calculate matrix
$(E+D_{0K}R_{0}(\lambda))^{-1}$ only once and then we need only $\sim
(2M+2N-2)^{2}+MN\log MN$ arithmetic operations for each right hand
side.  Asymptotically,  this complexity is the same as that for the
periodic Laplacian.

Note that in contrast with the one-dimensional case,    the direct (i.e.,
non-iterational) methods for calculation of $R_{K}(\lambda)$ exist only
for exceptional  extension operators even from family (\ref{bouloc}),
see \cite{Bah}. It makes Example 3.2 important in practical applications.

\begin{example}
{\bf The Laplacian in the two-dimensional rectangle
with a hole.} Let $\mathcal M$,  $\Delta$ and $\Omega$ be the same as in
Example 3.2 and $p$ an inner point of $\Omega$.  Let
$\Omega_{p}=\Omega\setminus\{p\}$.  It is clear that $p\in
b_{\Delta}\Omega_{p}$ and $ b_{\Delta}\Omega_{p}=\{p\}\cup
b_{\Delta}\Omega$.  We consider the extension operator $K_{p}$ of the
form (\ref{bouloc}) and suppose in addition that
\[
(K_{p}f)(p) = \sum_{\{\varepsilon:
|\varepsilon|=1\}}\alpha(\varepsilon)f(p+\varepsilon),  \qquad
\alpha(\varepsilon)\in {\mathbb C}.
\]
Note that in this formula $p+\varepsilon \in \Omega_{p}$ for
all $\varepsilon$  due to our hypotheses.

It is easy to see that operator $\Delta_{K_{p}}$ is exactly a rank~$1$
perturbation of $\left.\Delta_{K}\right|_{C(\Omega_{p})}$,  where we
consider the space $C(\Omega_{p})$ as a subspace in $C(\Omega)$
consisting of functions $f$ such that $f(p)=0$.  So the resolvent of
$\Delta_{K_{p}}$ can be calculated with the same efficiency as that of
$\Delta_{K}$!  This is indeed remarkable,  because one can consider
operator $\Delta_{K_{p}}$ as the difference approximation of the
differential Schr\"{o}dinger operator with point-wise potential~\cite{Alb},
and we see that the difference case can be investigated
with the help of the introduced finite dimensional analog of the Krein
formula in the same manner as differential operators with point-wise
potentials are investigated by means of the ``actual'' Krein formula.

It is clear that in the same way one can construct resolvents for
Laplacian in rectangle with more complicated defects (like holes containing
more than one point,  cuts,  etc).  Our approach is efficient provided
$\#(b_{A}\Omega)\ll\# (\Omega)$ and we know an effective
algorithm for calculating resolvent of at least one extension.
\end{example}

\renewcommand{\footnoterule}{\vspace*{3pt}%
\noindent
\rule{.4\columnwidth}{0.4pt}\vspace*{6pt}}

\begin{example}
{\bf The point-wise potentials in one-dimensional
case and convergence.} The aim of this example is to demonstrate that
in simplest case application of formula (\ref{Krein}) to the
difference approximation of differential operator leads to the
expression for resolvent which term-by-term converges to one obtained
by applying the ``actual'' Krein's formula to initial differential
operator.

Let $A$ be the Laplace operator ${-\frac{d^2}{dx^2}}$ in
$L_2([0,  2\pi])$ with periodic boundary conditions. It is evident that its
resolvent is of form
\begin{equation}\label{PL}
(R_A(\lambda)\varphi)(x)=\frac{1}{2\pi}\sum_{m=-\infty}^\infty
                \varphi_m{\rm e}^{{\rm i}m x}\frac{1}{m^2-\lambda},
\end{equation}
where
\begin{equation}\label{FR}
\varphi_m=\int_0^{2\pi}\varphi(x){\rm e}^{-{\rm i}m x}dx.
\end{equation}

Following Krein, consider the one-parametric family of self-adjoint
extensions of the restriction of $A$ onto the space of smooth
functions vanishes in the neighborhood of the endpoints of the
interval $[0, 2\pi]$, such that the resolvents of operators from the
family are of the form
\begin{gather}
   (R_{A_\mu}(\lambda)\varphi)(x)=
    \frac{1}{2\pi}\sum_{m=-\infty}^\infty
                \varphi_m{\rm e}^{{\rm i}m x}\frac{1}{m^2-\lambda}\nonumber\\
\qquad {}- \mu\frac{\displaystyle \left(\frac{1}{2\pi}
    \sum\limits_{m=-\infty}^\infty\frac{\varphi_m}{m^2-\lambda}\right)\left(
    \frac{1}{2\pi}\sum\limits_{m=-\infty}^\infty\frac{{\rm e}^{{\rm i}m
    x}}{m^2-\lambda}\right)}
    {\displaystyle 1+\frac{\mu}{2\pi}\sum\limits_{m=-\infty}^\infty\frac{1}{m^2-\lambda}},\label{LwDel}
\end{gather}
where $\mu$ is a parameter of family.  (Note that all series in this
expression converge either in $L_2([0, 2\pi])$ or in $\mathbb{C}$ when
$\varphi \in L_2([0, 2\pi]$).)  It is well-known (see, e.g.,
\cite{Alb}, where a number of similar examples are considered), that
for each real $\mu$ the operator $R_{A_\mu}$ is indeed the resolvent
of a self-adjoint operator $A_\mu$ in $L_2([0, 2\pi])$.  This $A_\mu$
is usually called the {\it Schr\"odinger operator with
$\delta$-potential} (parameter $\mu$ plays the role of a coupling
constant)\footnote{Of course, formula (\ref{LwDel}) does not give {\it
all} possible extension of symmetric operator considered, but the
family described suffices for our goals.}.

Let now $M$ be a positive integer and let operator $A_M$ in
$L_2(\{0, \dots,  2M-1\}])$ be of the form
\begin{equation}\label{ApLap}
  (A_M f)_j=-\frac{1}{h^2}(f_{j-1}-2f_j+f_{j+1}), \qquad j\in
  \{0, \dots,  2M-1\},
\end{equation}
where $h=\pi/M$ and the ``exterior" values of $f$ are defined by
``periodic boundary conditions" (\ref{PBC}).  It is easy to see that
the resolvent of operator $A_M$ is, due to relation (\ref{res0}), of
the form
\begin{equation}\label{DRes}
 ( R_{A_M}(\lambda)\varphi)_j=\frac{1}{2\pi}\sum_{m=-M+1}^M
                \widehat{\varphi}_m{\rm e}^{{\rm i}h m j}
                \frac{1}{\displaystyle \frac{4}{h^2}\sin^2\frac{h m}{2}-\lambda},
\end{equation}
where
\begin{equation}\label{DFR}
\widehat{\varphi}_m=h\sum_{m=0}^{2M-1}\varphi_j{\rm e}^{-{\rm i}h m j}.
\end{equation}
Observe that relations (\ref{PL}), (\ref{FR}) and (\ref{DRes}),
(\ref{DFR}) are of similar form.  Moreover, setting $T_h:C([0,
2\pi])\to L_2(\{0, \dots, 2M-1\})$ as
\begin{equation}\label{Prj}
  (T_hf)_j=f(h j),  \qquad j\in\{0, \dots, 2M-1\},
\end{equation}
one can see that that the relations
\begin{equation}\label{Conv1}
  \lim\limits_{M\to\infty}\|T_hAf-A_MT_hf\|_h=0, \qquad
  \lim\limits_{M\to\infty}\|T_hR_A(\lambda)f-R_{A_M}(\lambda)T_hf\|_h=0
\end{equation}
hold for every sufficiently smooth periodic function $f\in L_2([0, 2\pi])$ if the
norm in $L_2(\{0, \dots$, $2M-1\})$ is
\[
\|\varphi\|_h^2=h\sum_{j=0}^{2M-1}|\varphi_j|^2.
\]
This means exactly that the family of finite dimensional operators
$A_M, \ M\in \mathbb{N}$, {\it approximates} the operator $A$
\cite{Bah,Gor}, or, in another words, $A_M$ tends to $A$ when
$M\longrightarrow\infty$.

Let now $\widehat{\delta}^M_0$ be the operator in $L_2(\{0, \dots,
2M-1\})$ given by the formula
\begin{equation}\label{dDelta}
(\widehat{\delta}^M_0f)_j=\frac{1}{h}f_0\delta_{0j}.
\end{equation}
It is easy to make use of (\ref{Krein}) in order to
calculate the resolvent of
$A_{M\mu}\equiv A_M+\mu\widehat{\delta}^M_0$ (cf.  also with Example 2.1 and relation
(\ref{Ex1})). In this way we obtain an expression for the
resolvent of $A_{M\mu}$:
\begin{gather}     (R_{A_{M\mu}}(\lambda)\varphi)_j=
    \frac{1}{2\pi}\sum\limits_{m=-M+1}^M
                \varphi_m{\rm e}^{{\rm i}h m j}\frac{1}{\displaystyle \frac{4}{h^2}\sin^2\frac{h
                m}{2}-\lambda}\nonumber\\
\qquad {}-\mu\frac{\displaystyle \left(\frac{1}{2\pi}
    \sum\limits_{m=-M+1}^M\frac{\varphi_m}{\frac{4}{h^2}\sin^2\frac{h
    m}{2}-\lambda}\right)\left(
    \frac{1}{2\pi}\sum\limits_{m=-M+1}^M\frac{{\rm e}^{{\rm i}h m j}}
    {\frac{4}{h^2}\sin^2\frac{h m}{2}-\lambda}\right)}
    {\displaystyle 1+\frac{\mu}{2\pi}\sum\limits_{m=-M+1}^M\frac{1}{\frac{4}{h^2}\sin^2\frac{h
    m}{2}-\lambda}}.\label{dLwDel}
\end{gather}
We compare now relations (\ref{LwDel}) and (\ref{dLwDel}).  It is
easy to see that for resolvents $R_{A_\mu}$ and $R_{A_{M\mu}}$ a
relation like (\ref{Conv1}) holds.  Moreover, one can see also that
{\it each term} in the left hand side of (\ref{dLwDel}) converges to
the corresponding term in (\ref{LwDel}).  Hence, one can assert that
in this sense the finite dimensional Krein formula converges to the
``natural'' Krein formula for the resolvent $R_{A_\mu}$.  This fact gives
an additional argument in favor of the name ``finite dimensional
analog of Krein formula'' for relation (\ref{Krein}).
\end{example}

\section{Concluding remarks}\label{CR}

The finite dimensional analog of Krein's formula proved to
be a useful instrument for investigation of difference equations both
analytically and numerically.  Moreover, it gives us a new approach to
study {\it differential} problems (and, more generally, other
``continuous'' extensions) by reducing them to the corresponding
difference (or, more generally, other finite dimensional)
approximations.  In this connection it is interesting that, in
contrast with the ``actual'' Krein formula, our algebraic relation
does not require operators involved to be Hermitian.

The method proposed for solving difference boundary value
problem is applicable to a wide class of equations, in particular, in
the case of complicated multi-point boundary conditions for
one-dimensional equations, for the rectangular two-dimensional domains
with cuts and some other ``small" defects, for some cases of variable
coefficient of difference operators, etc.  It is clear that in every
specific case one needs to adapt the general algorithm described in
Sections~2, 3, but this general scheme is, nevertheless, useful for
construction of particular numerical procedures.

The described method for solving of boundary value problems
was successfully used in~\cite {GL}.

A similar approach exists also for constructing other than
resolvent functions of difference operators.  This is needed,  e.g.,  in
{\it initial boundary value problems},  see \cite {Nef}.  Certain
moments of the method proposed for solving difference boundary value
problems were announced in \cite{OkSh}.

\subsection*{Acknowledgements}
I am thankful to  M~Antonets,  E~Gordon and I~Nefedov
for helpful discussions and D~Leites for hospitality and support.

\label{sheresh-lastpage}

\begin{thebibliography}{99}
\small

\bibitem{AG}
Akhiezer N~I and  Glazman I~N, Theory of Linear Operators in
Hilbert Space; Second revised and augmented edition, Nauka~- Moscow,
1966  (in Russian); Third edition, corrected and augmented.
Vishcha Shkola~- Kharkov, Vol.~I, 1977, Vol.~II, 1978 (in Russian);
 Translated from the Russian and
with a preface by Merlynd Nestell.  Reprint of the 1961 and 1963
translations.  Two volumes bound as one.  Dover Publications, Inc.~-
New York, 1993.

\bibitem{Graph}
Gerasimenko N~I and Pavlov B~S,  Scattering Problems on
Noncompact Graphs,  {\it Teoret. Mat Fiz.}  {\bf 74}  (1988), 345--359 (traslation in
{\it Theor. and Math. Phys.}  {\bf 74}  (1988), 230--240).

\bibitem{Graph2}
Kostrykin V and Shrader R,  Kirchoff Rule for Quantum Wires,
{\it J. Phys. A: Math. Gen.} {\bf 32}  (1999),  595--630.

\bibitem{Alb}
Albeverio S,   Gesztesy F,   H{\o}egh-Krohn   R and    Holden~H,
Solvable Models in Quantum Mechanics,  Texts and Monographs in
Physics,  Springer-Verlag~-   New York~- Berlin,   1988.

\bibitem{Gor}
Gordon E I, Nonstandard Methods in Commutative Harmonic Analysis,
Translations of Mathematical Monographs, Vol.~164,
Providence,  R.I.,  American Mathematical Society,  1997.

\bibitem{AlbKur}
Albeverio S and Kurasov P (Editors), Singular Perturbations of
Differential Operators, London Mathematical Society Lecture Notes,
Vol.~ 271, Cambridge Univ.  Press~- Cambridge, 2000.

\bibitem{KurKur}
Kurasov P and Kuroda~T, Krein's Formula and Perturbation Theory,
Preprint Nr.~6, 2000, Dept.  of Math., Univ.  of Stocholm
(http://www.matematik.su.se).

\bibitem{Pav}
Pavlov B~S, Theory of Extensions and Exact Solable Models, {\it  Uspekhi
Matematicheskih Nauk  (Russian Mathematical Survays)} {\bf 42}, Nr.~6 (1987),
99--131 (in Russian).

\bibitem{Ant}
Antonets M A, Initial-Boundary  Value problems  for Evolution
Equation with Transmission Condition  on   an  Unbounded   Surface,
{\it Russian Acad. Sci. Dokl. Math.} {\bf 48}, Nr.~2 (1994), 286--290
({\it Ross. Acad. Nauk Dokl.} {\bf 332}, Nr.~3 (1993), 277--279).

\bibitem{Bah}
Bakhvalov N S, Zhidkov N P and Kobelkov G~M, Numerical
Methods, Nauka~- Moscow, 1987  (in Russian).

\bibitem{GL}
Vysheslavtsev P~P, Kurin V~V, Nefedov I~M, Shereshevsky I~A and
Andronov A~A,  Modelling of the Resistance State of Superconducting
Layers in the Magnetic Field on the Basis of the Ginzburg--Landau
Nonstationary Equation, {\it Izvestija VUZ'ov, Radiofizika} {\bf 40} (1997),
213--231 (in Russian).

\bibitem{Nef}
Nefedov I M and Shereshevskii I A, On  Solving of the Difference
Initial Boundary Value Problems by the Operator Exponential Method,
{J. Nonlin. Math. Phys.} {\bf 8},  Nr.~3 (2001),  313--324.

\bibitem{OkSh}
Okomelkova I A and Shereshevskii I A, Fast Method of Resolvent
Calculation for Difference Boundary Problems, {\it Mat.  Model.} {\bf 7}, Nr.~5 (1995), 89
(in Russian).

\bibitem{P}
Prasolov V~V, Problems and Theorems in Linear Algebra,
Translations of Mathematical Monographs, Vol.~134,  American Mathematical
Society, Providence, RI, 1994.

\end{thebibliography}
\end{document}